\documentclass[12pt,epsbox]{article}
\usepackage{latexsym}
\usepackage{psfig,amssymb,epsf}

\makeatletter
\@addtoreset{equation}{section}
\renewcommand{\theequation}{\thesection.\arabic{equation}}

\newcommand{\bR}{{\bf R}}

\newcommand{\cA}{{\cal A}}
\newcommand{\cJ}{{\cal J}}
\newcommand{\cK}{{\cal K}}

\newcommand{\cM}{{\cal M}}

\newcommand{\Kt}{{\widetilde K}}

\newcommand{\cJt}{\widetilde{\cal J}}

\newcommand{\nn}{\nonumber \\}

\newcommand{\be}{\begin{equation}} \newcommand{\ee}{\end{equation}}
\newcommand{\bea}{\begin{eqnarray}} \newcommand{\eea}{\end{eqnarray}}

\font\zfont = cmss10 
\newcommand{\ZZ}{\hbox{\zfont Z\kern-.4emZ}}

\ifx\TwoupWrites\UnDeFiNeD\else\target{\magstepminus1}{11.3in}{8.27in}
\source{\magstep0}{7.5in}{11.69in}\fi
\newfont{\fourteencp}{cmcsc10 scaled\magstep2}
\newfont{\titlefont}{cmbx10 scaled\magstep3}
\newfont{\authorfont}{cmcsc10 scaled\magstep1}
\newfont{\fourteenmib}{cmmib10 scaled\magstep2}
\skewchar\fourteenmib='177
\newfont{\elevenmib}{cmmib10 scaled\magstephalf}
\skewchar\elevenmib='177
\makeatletter
\newcommand\nonsequentialeqnum{
\@addtoreset{equation}{section}
\def\theequation{\arabic{section}.\arabic{equation}}}
\newif\ifp@bblock \p@bblocktrue
\newcommand\nopubblock{\p@bblockfalse}
\newcommand\topspace{\hrule height 0pt depth 0pt \vskip}
\newcommand\p@bblock{\begingroup \tabskip=\hsize minus \hsize
\baselineskip=1.5\ht\strutbox \topspace-2\baselineskip
\halign to\hsize{\strut ##\hfil\tabskip=0pt\crcr
\the\Pubnum\crcr\the\date\crcr}\endgroup}

\renewcommand\titlepage{\ifx\TwoupWrites\UnDeFiNeD\null
\vspace{-1.7cm}\fi
\vskip0.6cm
\ifp@bblock\p@bblock \else\hrule height 0pt \relax \fi}
\makeatother
\newtoks\date
\newtoks\Pubnum
\newtoks\pubnum
\newcommand{\frontpageskip}{\vspace{12pt plus .5fil minus 2pt}}
\renewcommand{\title}[1]{\frontpageskip
\begin{center}{\titlefont #1}\end{center}\par}
\renewcommand{\author}[1]{\frontpageskip\par\begin{center}
{\authorfont #1}\end{center}
\nobreak
}

\renewcommand{\thanks}[1]{\footnote{#1}}
\renewcommand{\abstract}{\par\frontpageskip\centerline{
\fourteencp Abstract}
\vspace{8pt plus 3pt minus 3pt}}

\begin{document}

\begin{titlepage}
\hfill
\vbox{
    \halign{#\hfil         \cr
           TAUP-2726-03 \cr
           hep-th/0306112  \cr
           } 
      }  
\vspace*{20mm}
\begin{center}
{\Large {\bf  D-branes in $N=2$ Strings}\\} 
\vspace*{15mm}
{\sc Dan Gl\"uck}
\footnote{e-mail: {\tt gluckdan@post.tau.ac.il}}
{\sc Yaron Oz} 
\footnote{e-mail: {\tt yaronoz@post.tau.ac.il, Yaron.Oz@cern.ch}}
and {\sc Tadakatsu Sakai}
\footnote{e-mail: {\tt tsakai@post.tau.ac.il}}

\vspace*{1cm} 
{\it {$^{a}$ Raymond and Beverly Sackler Faculty of Exact Sciences\\
School of Physics and Astronomy\\
Tel-Aviv University , Ramat-Aviv 69978, Israel}}\\

\end{center}

\begin{abstract}
We study various aspects of D-branes in the two families  of closed
$N=2$
strings denoted by $\alpha$ and $\beta$ in hep-th/0211147.
We consider two types of $N=2$ boundary conditions, A-type and B-type.
We analyse the D-branes geometry.
We compute open and closed string scattering amplitudes in the
presence
of the D-branes and discuss 
the results. We find that, except the space filling D-branes,
the B-type D-branes decouple from the bulk. The  A-type D-branes
exhibit inconsistency.        
We construct the D-branes effective worldvolume theories.
They are given 
by a dimensional reduction of self-dual Yang-Mills
theory in four dimensions.
We construct the D-branes gravity backgrounds.
Finally, we discuss possible $N=2$ open/closed string dualities.

\end{abstract}
\vskip .7cm

June 2003

\end{titlepage}

\setcounter{footnote}{0}

\newpage

\section{Introduction}

Closed $N=2$ strings \cite{ademollo;76, Gates:1988tn} possess local $N=2$
supersymmetry on the string worldsheet.
Critical $N=2$ strings have a four-dimensional target space.
The supersymmetric structure implies that the target space
has a complex structure. Therefore it must be of signature $(4,0)$
or $(2,2)$.
In $(4,0)$ signature there are no propagating degrees of freedom
in the $N=2$ string spectrum.
In $(2,2)$ signature there are only massless scalars  in the
spectrum and
the infinite tower of massive excitations of the string is absent.

In \cite{oz;n2} the
$N=2$
closed strings have been divided into two families denoted by
$\alpha$ and $\beta$.
Consider $N=2$ strings in a flat background.
In order to construct the $N=2$ string we need to gauge the $N=2$ 
superconformal algebra (SCA) on the worldsheet. 
More precisely we have two copies of the $N=2$
algebra to consider: the left and right sectors. 
The free field
representation of the (left) $N=2$ SCA
takes the form
\begin{eqnarray}
&&T=
-{1\over 2}\eta_{IJ}\partial x^I\partial x^J-{1\over 4}\,\eta_{IJ}
\left( \partial\psi^I\psi^J+\partial\psi^J\psi^I\right),\nn
&&J={i\over 2}\cJ_{IJ}^L\psi^I\psi^J,\nn
&&G^{\pm}={i\over 2}\left(
\eta_{IJ}\pm i\cJ_{IJ}^L\right)\psi^I\partial x^J \ .
\label{sca}
\end{eqnarray}
Here $I,J=1,...,4$ denote the indices of the target space in a real
basis.
The metric is given by
$\eta_{IJ}={\rm diag}(-1,-1,+1,+1)$.
${\cal J}^L_{IJ}$ is a K\"ahler form related to the complex structure
$\cJ^K_{~J}$
by $\cJ_{IJ}=\eta_{IK}\cJ^K_{~J}$, and 
the index $L$ refers to the left sector.
Similarly, 
we have the SCA generators in the right sector with a
complex structure ${\cal J}^R$.
The $N=2$ string denoted by $\beta$-string
in \cite{oz;n2} is defined by having
the same complex structure in the left and right sectors
${\cal J}^L={\cal J}^R$.
The $N=2$ string denoted as
$\alpha$-string in  \cite{oz;n2}
has different (inequivalent) complex structures in the left and
right sectors. 

The $\beta$- and $\alpha$-strings define two families
of $N=2$ strings related by T-duality
\cite{oz;n2}.
In $(2,2)$ signature both families have one scalar in the spectrum.
The effective action of the  $\beta$-string scalar has been computed in
\cite{ov}, which suggested its interpretation as a deformation of the
target space  K\"ahler potential. 
The effective action of the scalar
in the $\alpha$-string has been computed in 
\cite{gos}. It was found that the 
$\alpha$-string scalar is free and that the 
dynamics is that of a self-dual curvature with torsion considered in
\cite{Hull}.

In this paper we study the D-branes in both families
of $N=2$ strings.
We will mainly consider the target space with $(2,2)$ signature.
The extension to $(4,0)$ signature is straightforward.
We will consider two types of $N=2$ boundary conditions, A-type and
B-type
\cite{ooy,oz;n2}.
We will compute open and closed string scattering amplitudes in the
presence
of the D-branes and discuss 
the results. We find that, except the space filling D-branes,
the B-type D-branes decouple from the bulk: 
the perturbative closed string scattering amplitudes off B-type
branes vanish.
The  A-type D-branes
exhibit an inconsistency: the scattering
amplitude of two closed string modes in the A-branes background  contains an 
infinite number of poles that correspond to massive excitations of both
open and closed string states.         
Such an observation was first made in \cite{js;2001}.

We construct the D-branes effective worldvolume theories.
They are given 
by a dimensional reduction of self-dual Yang-Mills
theory in four dimensions.
They correspond to  various integrable systems: 
Bogomolny equation \cite{bogomolny} for the three-branes, the Hitchin system \cite{hitchin}
for the two-branes, Nahm equations for the one-branes \cite{Nahm:1979yw}
and the 
ADHM equation \cite{adhm} for the zero-branes.
We construct the D-branes gravity backgrounds. 
Finally, we comment on $N=2$ open/closed string dualities.

The paper is organized as follows.
In section 2 we consider the A-type and B-type boundary conditions
on the $N=2$ algebra. We discuss the classification of $N=2$ D-branes
and their geometry.
In section 3 
we construct vertex operators and
compute open and closed string scattering amplitudes in the
presence
of the D-branes. 
In section 4 
we construct the D-branes effective worldvolume theories
as a dimensional reduction of self-dual Yang-Mills
theory in four dimensions and see 
the correspondence to  integrable systems.
In section 5 
we construct the D-branes gravity backgrounds. 
In section 6
we discuss possible $N=2$ open/closed string dualities.

\section{$N=2$ Boundary Conditions}

In this section we consider the A-type and B-type boundary conditions
on the $N=2$ algebra and the corresponding $N=2$ D-branes.
We will expand the discussion of section 7 in \cite{oz;n2}.

\subsection{Closed $N=2$ Strings}

In the following we will review some aspects
of the complex structures on $\bR^{2,2}$ that are relevant to
the generators of $N=2$ SCA.
In the real basis $x^I=(x^1,x^2,x^3,x^4)$, the metric is given by
$\eta_{IJ}={\rm diag}(-1,-1,+1,+1)$.
We define a complex structure
\begin{equation}
{\cal J}^I_{~J}=
\left(
\begin{array}{cc}
i\sigma_2 & 0 \\
0         & i\sigma_2 
\end{array}
\right) \ .
\end{equation}
In the complex basis
\begin{equation}
z^1={x^1+ix^2\over\sqrt{2}},~~z^2={x^3+ix^4\over\sqrt{2}} \ ,
\label{basis}
\end{equation}
the metric reads $\eta_{i\bar{j}}={\rm diag}(-1,+1)$, $i,\bar{j} =1,2$.
In this basis, the complex structure ${\cal J}^I_{~J}$ is diagonal:
\begin{eqnarray}
&&\cJ(z^1)=-i z^1, ~~\cJ(\bar{z}^{\bar 1})=+i \bar{z}^{\bar 1} \ , \nn
&&\cJ(z^2)=-i z^2, ~~\cJ(\bar{z}^{\bar 2})=+i \bar{z}^{\bar 2} \ .
\end{eqnarray}
The K\"ahler form $\cJ_{IJ}=\eta_{IK}\cJ^K_{~J}$ is given in the real
basis
by
\begin{equation}
\cJ_{IJ}=
\left(
\begin{array}{cc}
-i\sigma_2 & 0 \\
0         & i\sigma_2 
\end{array}
\right) \ .
\end{equation}
For later reference, we define the quadratic form in momenta
\begin{equation}
\label{cij}
k_i^I\cJ_{IJ}k_j^J=ic_{ij} \ ,
\end{equation}
where in the complex basis
\begin{equation}
\label{cijn}
c_{ij} = k_i\cdot\bar{k}_j-\bar{k}_i\cdot k_j \ ,
\end{equation}
and $ k_i\cdot\bar{k}_j \equiv \eta_{m\bar{n}} k_i^m\bar{k}_j^{\bar{n}}$.
For the on-shell momenta $k_i,~i=1,2,3,4$ with 
$k_i^2=0,~k_1+k_2+k_3+k_4=0$,
$c_{ij}$ obey the identities \cite{ov}
\begin{eqnarray}
{c_{12}c_{34}\over s} + {c_{23}c_{41}\over t} = u,~~
{c_{21}c_{34}\over s} + {c_{13}c_{42}\over u} = t,~~
{c_{13}c_{24}\over u} + {c_{32}c_{41}\over t} = s \ ,
\label{id;c}
\end{eqnarray}
where $s=-k_1\cdot k_2 \equiv -(k_1\cdot\bar{k}_2+\bar{k}_1\cdot k_2),
~t=-k_2\cdot k_3,~u=-k_1\cdot k_3$.

We will also need a second complex (and  K\"ahler) structure,
which in the real basis take the form
\begin{equation}
\widetilde{{\cal J}}^I_{~J}=
\left(
\begin{array}{cc}
i\sigma_2 & 0 \\
0         & -i\sigma_2
\end{array}
\right),\quad
\widetilde{\cJ}_{IJ}=\eta_{IK}\widetilde{\cJ}^K_{~J}=
\left(
\begin{array}{cc}
-i\sigma_2 & 0 \\
0         & -i\sigma_2
\end{array}
\right) \ .
\end{equation}
In the complex basis, the complex structure is given by
\begin{eqnarray}
&&\cJt(z^1)=-i z^1, ~~\cJt(\bar{z}^{\bar 1})=+i \bar{z}^{\bar 1} \ , \nn
&&\cJt(z^2)=+i z^2, ~~\cJt(\bar{z}^{\bar 2})=-i \bar{z}^{\bar 2} \ .
\end{eqnarray}
The K\"ahler form reads
\begin{equation}
\cJt_{i\bar{j}}=-i{\bf 1},~~\cJt_{\bar{i}j}=+i{\bf 1} \ .
\end{equation}

The $\beta$-string in $\bR^{2,2}$ is defined by the choice
\begin{equation}
\cJ^L_{IJ}=\cJ^R_{IJ}=\cJ_{IJ} \ ,
\end{equation}
in  (\ref{sca}), while 
for the $\alpha$-string 
\begin{equation}
\cJ^L_{IJ}=\cJ_{IJ},\quad \cJ^R_{IJ}=\widetilde{\cJ}_{IJ} \ .
\end{equation}
The two $N=2$ strings are related 
by a T-duality along a spatial direction.

\subsection{A-type and B-type Boundary Conditions}

We consider the two types of $N=2$ boundary conditions:
the A-type and B-type.
In the closed string notation 
the boundary conditions read  \cite{ooy}:
\begin{itemize}
\item {\bf A-type}:
\begin{eqnarray}
\left( L_n-\widetilde{L}_{-n}\right)|~\rangle\!\rangle_A
=\left( J_n-\widetilde{J}_{-n}\right)|~\rangle\!\rangle_A
=\left( G_n^{\pm}-i\widetilde{G}_{-n}^{\mp}\right)|~\rangle\!\rangle_A
=0 \ , 
\label{Abc}
\end{eqnarray}

\item {\bf B-type}:
\begin{eqnarray}
\left( L_n-\widetilde{L}_{-n}\right)|~\rangle\!\rangle_B
=\left( J_n+\widetilde{J}_{-n}\right)|~\rangle\!\rangle_B
=\left( G_n^{\pm}-i\widetilde{G}_{-n}^{\pm}\right)|~\rangle\!\rangle_B
=0 \ .
\label{Bbc}
\end{eqnarray}
\end{itemize}
In order to solve these equations we use the oscillator modes:
\begin{equation}
\left( \alpha_n^I-U^I_{~J}\widetilde{\alpha}_{-n}^J\right)
|~\rangle\!\rangle_{A,B}
=\left( \psi_n^I-iV^I_{~J}\widetilde{\psi}_{-n}^J\right)
|~\rangle\!\rangle_{A,B}=0 \ .
\label{UV}
\end{equation}
The consistency condition that these operators (anti-)commute with each
other implies that 
$U$ and $V$ are elements of 
$O(2,2)$. Using (\ref{UV}) in (\ref{Abc}) and (\ref{Bbc})
and the free field
representation
of the $N=2$ algebra (\ref{sca}) we get
\begin{itemize}
\item {\bf A-type}:
\begin{equation}
U=V,\quad (U^T\cJ^L U)_{IJ}=-\cJ^R_{IJ} \ ,
\end{equation}

\item {\bf B-type}:
\begin{equation}
U=V,\quad (U^T\cJ^L U)_{IJ}=+\cJ^R_{IJ} \ .
\end{equation}
\end{itemize}

For the $\beta$-string $\cJ^L_{IJ}=\cJ^R_{IJ}=\cJ_{IJ}$ and
we get 
\begin{itemize}
\item {\bf A-type}:
\begin{equation}
U=V=\left(
\begin{array}{cc}
\sigma_3 & 0 \nn
0 & \sigma_3
\end{array}
\right),
\end{equation}
defining  a (1+1)-brane.

\item {\bf B-type}:
\begin{equation}
U=V=\left(
\begin{array}{cc}
\pm 1  & 0 \nn
0 & \pm 1
\end{array}
\right),
\end{equation}
defining  a (2+2), (2+0), (0+2) and (0+0)-brane depending on the number
of $+1$ eigenvalues.

\end{itemize}

For the $\alpha$-string 
$\cJ^L_{IJ}=\cJ_{IJ},\quad \cJ^R_{IJ}=\widetilde{\cJ}_{IJ}$ and
we get 
\begin{itemize}
\item {\bf A-type}:
\begin{equation}
U=V=\left(
\begin{array}{cc}
\sigma_3 & 0 \nn
0 & \pm 1
\end{array}
\right),
\end{equation}
defining a (1+2) and (1+0)-brane.

\item {\bf B-type}:
\begin{equation}
U=V=\left(
\begin{array}{cc}
\pm 1  & 0 \nn
0 & \sigma_3
\end{array}
\right),
\label{UV;B}
\end{equation}
defining a (2+1) and (0+1)-brane.
\end{itemize}

\subsection{Geometry of D-branes}

We start by considering $N=2$ strings in a flat  $\bR^{2,2}$ background.
The B-type boundary conditions in the $\beta$-string
correspond to even-dimensional D-branes.
The analysis of \cite{ooy} implies that
the worldvolumes are holomorphic (K\"ahler) submanifolds of $\bR^{2,2}$ 
with the complex structure $\cJ$.
The space-time filling (2+2)-brane is the target space $\bR^{2,2}$
itself. If we use the complex coordinates
on  $\bR^{2,2}$ $z_1, z_2$ in (\ref{basis}), then the (2+0) and (0+2)-branes 
are the holomorphic 2-cycles 
$[z_1, z_2=0]$ and $[z_1=0, z_2]$, respectively.
The (0+0)-brane is a point in $\bR^{2,2}$.

The A-type boundary conditions in the $\beta$-string
correspond to a (1+1)-brane. The worldvolume cycle is
a special Lagrangian
submanifold of $\bR^{2,2}$ \cite{ooy}. Denote the (1+1)-brane worldvolume
coordinates by $x^1,x^3$. The pull-backs of the
holomorphic 2-form $\Omega$ and K\"ahler 2-form $\cK$ read
\begin{equation}
\Omega \big|_{\rm worldvolume}=dx^1\wedge dx^3 \ ,~~
\cK \big|_{\rm worldvolume}=0 \ .
\end{equation}

The A and B-type boundary conditions
in the $\alpha$-string
correspond to odd-dimensional D-branes. 
We consider  $\bR^{2,2}$ as a product of two one-dimensional  K\"ahler
manifolds $\cM_1$ and $\cM_2$.
Consider the (2+1)-brane. 
The worldvolume is a product of 
a two-dimensional cycle, say  $M_1 = \cM_1$ parametrized  by $ x^1,x^2$ and 
the one-dimensional cycle $M_2 \in \cM_2$ parametrized  by $x^3$.
We will now show that $M_1$ is a holomorphic 2-cycle and $M_2$
is a Lagrangian submanifold with respect to the complex
structure $\cJ$ or $\cJt$.
We will follow the discussion
in \cite{lz}.

We write the matrix $U$ (\ref{UV;B}) as 
\begin{equation}
U=U^++U^- \ ,
\label{deco}
\end{equation}
where
\begin{equation}~
U^+=\left(
\begin{array}{cc}
1 & 0 \\
0 & 0
\end{array}
\right) \ ,
~~~~
U^-=\left(
\begin{array}{cc}
0 & 0 \\
0 & \sigma_3
\end{array}
\right) \ .
\end{equation}
$U^+$ defines the cycle $M_1$ and $U^-$ defines the cycle
$M_2$. 
It is easy to see that the following relations hold
\begin{equation}
U^+\cJ=+\cJ U^+ \ , ~~
U^-\cJ=-\cJ U^- \ , 
\end{equation}
and
\begin{equation}
U^+\cJt=+\cJt U^+ \ , ~~
U^-\cJt=-\cJt U^- \ . 
\end{equation}
These imply that
$M_1$ is a holomorphic 2-cycle and $M_2$
is a Lagrangian submanifold with respect to the complex
structures $\cJ$ or $\cJt$.
The $(0+1)$-brane is a point in $\cM_1$ times 
a Lagrangian submanifold with respect to the complex
structures $\cJ$ or $\cJt$ in $\cM_2$.

Consider next the D-branes in a curved background ${\cal M}$. 
The analysis of the D-branes in the $\beta$-string is done
in \cite{ooy} and
the worldvolumes are holomorphic (sub)manifolds of ${\cal M}$.
Consider next the $\alpha$-string.
The target space geometry
of $\alpha$-string is given by a bi-Hermitian geometry with
two commuting complex structures $\cJ^{\pm}$ that are covariantly
constant with respect to the affine connection with a torsion  
$\nabla^{\pm}\cJ^{\pm}=0$ with 
$\Gamma^{\pm\rho}{}_{\mu\nu}=\Gamma^{\rho}{}_{\mu\nu}\mp
H^{\rho}_{\mu\nu}$ \cite{GHR,Hull}.
One defines the projection operator \cite{lz}
\begin{equation}
\Pi_{\pm}={1\over 2}\,(I\pm\Pi) \ ,
\end{equation}
with
\begin{equation}
\Pi=\cJ^+\cJ^- \ , ~~~\Pi^2=1 \ .
\end{equation}
It defines a local product structure of the target space
geometry: ${\cal M}$ is locally ${\cal M}_1\times {\cal M}_2$
where $\cM_1$ and $\cM_2$ are  K\"ahler manifolds. 
The D-branes have a local product structure
as well.
The matrix $U$ can de decomposed as in (\ref{deco}) 
with
\begin{equation}
U^{\pm}=\Pi_{\pm}U\Pi_{\pm} \ .
\end{equation}
The boundary conditions require the 
the solution of two independent relations in  the K\"ahler geometry of
$\cM_1$ and $\cM_2$
\begin{eqnarray}
U^+\cJ_+&\!=\!&\eta \cJ_+U^+ \ , \nn
U^-\cJ_+&\!=\!&-\eta \cJ_+U^- \ ,
\end{eqnarray}
where $\eta=1,-1$ for the B- and A- type boundary conditions, respectively.
When $\eta=1$ the brane geometry is a local product of a holomorphic
cycle
in $\cM_1$ and a Lagrangian cycle in $\cM_2$
with respect to the complex structure $\cJ^+$. 
The structure is interchanged when  $\eta=-1$.

\section{$N=2$ Strings Scattering off D-branes}

In this section we compute string scattering
amplitudes in the presence of A-type and B-type branes.
We will see that in the presence of
A-branes the amplitudes exhibit infinite number of
massive poles, which do not correspond to $N=2$ string states.
The amplitudes in the presence of B-branes vanish and imply a
decoupling between the open and closed string states.
We note that $N=2$ strings scatterings in the presence of D-branes
 were studied in
\cite{js;2001}, before their A and B classification \cite{oz;n2}.
There will be some overlap between our computations
and those of \cite{js;2001}.

\subsection{The Vertex Operators}

We start by constructing the vertex operators of the closed
$N=2$ string 
scalar $\phi$ in the different pictures.
We consider both the $\alpha$ and $\beta$ strings.
We analyze first the matter part of the vertex operators. The
ghost part will be discussed later.

The scalar vertex operator of the matter
sector part reads in the $(-1,-1)$-picture  
\begin{equation}
V_L^{(-1,-1)}(z)=e^{ik\cdot X_L(z)},\quad
V_R^{(-1,-1)}(\bar{z})=e^{ik\cdot X_R(\bar{z})} \ .
\end{equation}
The picture-changing operators \cite{fms} are given by the world sheet
supercharges:
\begin{equation}
G_L^{\pm}(z)={i\over 2}\left(
\eta_{IJ}\pm i\cJ_{IJ}^L\right)\psi^I_L\partial X^J_L,\quad
G_R^{\pm}(\bar{z})={i\over 2}\left(
\eta_{IJ}\pm i\cJ_{IJ}^R\right)\psi^I_R\bar{\partial} X^J_R \ .
\end{equation}
Using the OPE's of the free fields 
\begin{equation}
X_L^I(z)X_L^J(w)\sim -\eta^{IJ}\log(z-w),~~
\psi_L^I(z)\psi_L^J(w)\sim -\frac{\eta^{IJ}}{z-w} \ ,
\end{equation}
one gets 
\begin{equation}
G_L^{\pm}(z)V_L^{(-1,-1)}(0)\sim {1\over z}\left(
V_L^{(0,-1)}(0),\,V_L^{(-1,0)}(0)\right) \ ,
\end{equation}
where
\begin{eqnarray}
V_L^{(0,-1)}(z)={1\over2}(kJ_-^L\psi_L(z))\,e^{ik\cdot X_L(z)},\quad
V_L^{(-1,0)}(z)={1\over2}(kJ_+^L\psi_L(z))\,e^{ik\cdot X_L(z)}  \ .
\end{eqnarray}
Here we used $J_{\pm}^L \equiv \eta\pm i\cJ^L$. 
Also
\begin{equation}
G_R^{\pm}(\bar{z})V_R^{(-1,-1)}(0)\sim {1\over \bar{z}}\left(
V_R^{(0,-1)}(0),\,V_R^{(-1,0)}(0)\right)  \ ,
\end{equation}
where
\begin{eqnarray}
V_R^{(0,-1)}(\bar{z})={1\over2}(kJ_-^R\psi_R(\bar{z}))\,e^{ik\cdot X_R(\bar{z})},
\quad
V_R^{(-1,0)}(\bar{z})=
{1\over2}(kJ_+^R\psi_R(\bar{z}))\,e^{ik\cdot X_R(\bar{z})} \ ,
\end{eqnarray}
with $J_{\pm}^R \equiv \eta\pm i\cJ^R$.
The vertex operator in the  $(0,0)$-picture is
\begin{equation}
G_L^-(z)V_L^{(0,-1)}(0)-G_L^+(z)V_L^{(-1,0)}(0)\sim
{1\over z}\,V_L^{(0,0)}(0) \ ,
\end{equation}
and 
\begin{equation}
V_L^{(0,0)}(z)=\left( -k{\cal J}^L\partial X_L+
{1\over 2}\left(kJ^L_+\psi_L\right)\left(kJ^L_-\psi_L\right)\right)
e^{ik\cdot X_L} \ .
\end{equation}
The right sector vertex operators take a similar form  
with $L\rightarrow R$.

Consider next the open string vertex operators.
We take the open strings on the upper-half plane
with the  real axis as a boundary.
The vertex operator of
the open $N=2$ scalar $\varphi$ \cite{mar;92}
reads in the $(-1,-1)$-picture 
\begin{equation}
V_o^{(-1,-1)}=e^{ik\cdot X_L+ik\cdot X_R} \ ,
\end{equation}
with the boundary conditions for $X$ and $\psi$ given by
\begin{equation}
X^{I}_L(z)=D^I_J X^J_R(\bar{z}),\quad
\psi_L^I(z)=D^I_J \psi_R^J(\bar{z})~~~{\rm at}~{\rm Im}z=0 \ ,
\end{equation}
where
\begin{equation}
D={\rm diag}(\pm\pm\pm\pm) \ .
\end{equation}
Here $+,-$ correspond to the Neumann(N), Dirichlet(D) boundary conditions,
respectively.

It is useful to use the doubling technique to implement the boundary
conditions. We define
\begin{eqnarray}
x^I(z)&\!=\!&\left(
\begin{array}{cc}
X_L^{I}(z), &{\rm Im}z\ge 0 \\
D^I_JX_R^{J}(z), &{\rm Im}z\le 0
\end{array}
\right. \nn
\Psi^I(z)&\!=\!&\left(
\begin{array}{cc}
\psi_L^{I}(z), &{\rm Im}z\ge 0 \\
D^I_J\psi_R^{J}(z), &{\rm Im}z\le 0
\end{array}
\right. 
\end{eqnarray}
with the OPE's given by
\begin{equation}
x^I(z)x^J(w)\sim -\eta^{IJ}\log(z-w),~~
\Psi^I(z)\Psi^J(w)\sim -\frac{\eta^{IJ}}{z-w} \ .
\end{equation}
In terms of these, one gets
\begin{equation}
V_o^{(-1,-1)}=e^{2ik\cdot x} \ .
\end{equation}
Here we used the fact that
$k$ has non-vanishing components only in the directions
parallel to a D-brane, i.e. the N directions.
Note that the matrix $D$ is equal to $U=V$, satisfying the
relations
\begin{equation}
DJ^R_{\pm}D=J^L_{{\mp}(\pm)} \ ,
\label{jljr}
\end{equation}
for the A-type (B-type) boundary conditions.

The picture-changing operators acting on the open string
vertex operator are given by
\begin{equation}
G^{\pm}={i\over 2}\Psi J_{\pm}^L\partial x \ .
\end{equation}
Thus, the open string vertex operators with various
superconformal ghost numbers are identical to the closed string
vertex operators in the left sector with $k\rightarrow 2k$:
\begin{eqnarray}
V_o^{(-1,0)}&\!=\!&\left(kJ_+^L\Psi\right)e^{2ik\cdot x},\nn
V_o^{(0,-1)}&\!=\!&\left(kJ_-^L\Psi\right)e^{2ik\cdot x},\nn
V_o^{(0,0)}&\!=\!&2\Big(-k{\cal J}^L\partial x
+\left( kJ_+^L\Psi\right)\left( kJ_-^L\Psi\right)
\Big)e^{2ik\cdot x} \ .
\end{eqnarray}
Note that both $\alpha$- and $\beta$-strings have
the same form of the open string vertex operators except for the momentum.
This fact will be important when we compute the scattering amplitudes
of open strings on D-branes.

In the following we  compute  string scattering amplitudes
in the presence of D-branes. 
The analysis is similar  to that of scatterings
in the presence of D-branes in $N=1$ superstrings
(For a review, see \cite{hk;review}).

\subsection{The Cylinder Amplitude}

Let us compute the one-loop amplitude of an $N=2$ open string that connects
two flat parallel Dp-branes
($p$ refers to the number of both space- and time-like directions).
This computation contains the information about  the force
between the D-branes \cite{Polchinski:1995mt}. As expected, we will 
obtain a non-vanishing force that is
mediated by the closed string scalar $\phi$. 
The
D-branes of $N=2$ strings do not posses RR charge and are not BPS objects.

The amplitude is defined by
\begin{equation}
\cA=\int_0^{\infty}{dt\over 2t}\int du\,d\bar{u}\,\,{\rm tr}_{\phi}\left(
e^{-2\pi tL_0}e^{2\pi i\theta J_0}\right) \ .
\end{equation}
$t$ is the modulus of a cylinder that corresponds to its radius.
$u=\phi+it\theta$ denotes the $N=2$ $U(1)$ moduli. $L_0,~J_0$ are the zero
modes of the Virasoro generators and the $N=2$ $U(1)$ current.
${\rm tr}_{\phi}$ is a summation over worldsheet fermionic modes
$\psi_n,
n\in Z+\phi$.
$L_0$ takes the form
\begin{equation}
L_0=\alpha^{\prime}p_{\parallel}^2+{r^2\over 4\pi^2\alpha^{\prime}}+N \ ,
\end{equation}
where $p_{\parallel}$ is the momentum along the D-branes worldvolume,
$r$ is the distance between the D-branes
and $N$ is the oscillation number.
As in the closed string case \cite{ov}, 
all the massive excitations do not contribute to the amplitude
due to a cancellation between the matter and ghosts sectors.

We obtain
\begin{eqnarray}
\cA&\!=\!&V_p\int_0^{\infty}dt 
\left( 8\pi^2\alpha^{\prime} t\right)^{-p/2}
\,e^{-{tr^2\over 2\pi\alpha^{\prime}}} \nn
&\!=\!& 2\pi V_p\left( 4\pi^2\alpha^{\prime}\right)^{1-p}G_{4-p}(r) \ .
\label{Amp1}
\end{eqnarray}
$V_p$ is the volume of the Dp-branes worldvolume and $G_{4-p}$ is the
harmonic function in a flat $\bR^{4-p}$ 
\begin{equation}
G_{4-p}(r)={1\over 4}\pi^{p-4\over 2}\,\Gamma\left( {2-p\over 2}\right)
r^{p-2} \ .
\end{equation}

The effective action of $\beta$-string is 
the Plebanski action \cite{ov} and that of $\alpha$-string is
a free real scalar action \cite{gos}. By adding an interaction
term to  Dp-branes  we have
\begin{equation}
S={1\over 2\kappa^2}\int d^4x\, (\partial\phi)^2
+\mu_p\int d^px\, \phi \ ,
\end{equation}
where we omitted the cubic interaction of the $\beta$-string effective action.
Note that $\phi$ has dimension of length squared, and
the target space metric obtained from $\phi$ is
dimensionless. Consequently $\kappa^2$ has a length dimension
$6$, and $\mu_p$ has a mass dimension $p+2$.
The action implies that the amplitude $\cA$ is 
\begin{equation}
\cA=\mu_p^2\kappa^2 V_p\, G_{4-p}(r) \ .
\label{Amp}
\end{equation}
Comparing (\ref{Amp1}) and (\ref{Amp}) we get
\begin{equation}
\mu_p^2={2\pi (4\pi^2\alpha^{\prime})^{1-p}\over \kappa^2} \ .
\end{equation}

Since the D-branes of $N=2$ strings are not BPS objects
there is a priori no reason to expect
that this result is not corrected by higher order terms.
However, this may still be the case 
since most of the higher order amplitudes of $N=2$ strings vanish.

\subsection{Closed Strings Scattering off D-branes}

Consider the scattering amplitudes of closed strings off
a D-brane. 
The computation involves the insertion of
 closed string vertex operators
on a disk with the boundary conditions 
\begin{equation}
X^{I}_L(z)=D^I_J X^J_R(\bar{z}),\quad
\psi_L^I(z)=D^I_J \psi_R^J(\bar{z}) \ ,
\end{equation}
at $z=\bar{z}$. 
We use the doubling technique to implement
the boundary conditions for the matter as well as the 
ghost sectors.

We will use in the sequel the following formula
for the ghost and bosonized superconformal ghosts correlators
\begin{eqnarray}
\langle c(z_1)c(z_2)c(z_3)\rangle&\!=\!&(z_1-z_2)(z_2-z_3)(z_1-z_3),
\nn
\left\langle e^{-\varphi^{\pm}(z_1)}e^{-\varphi^{\pm}(z_2)}\right\rangle
&\!=\!&{1\over z_1-z_2} \ .
\label{correlator;ghost}
\end{eqnarray}

\subsubsection{One Closed String}


Here we calculate directly the coupling of a D-brane to the closed
string 
scalar. It is given by the one-point function on the disk
of the closed string vertex operator
\begin{equation}
A_c={1\over 2\pi}\langle V_L^{(-1,-1)}(z)V_R^{(-1,-1)}(\bar{z})\rangle_{D_2}
\langle c(z)c(\bar{z})\rangle_{D_2}
\langle e^{-\varphi^+(z)} e^{-\varphi^+(\bar{z})}\rangle_{D_2}
\langle e^{-\varphi^-(z)} e^{-\varphi^-(\bar{z})}\rangle_{D_2}\nn \ .
\end{equation}
The $1\over 2\pi$ factor comes from the division by
the volume of the residual 
conformal Killing subgroup $U(1) \subset SL(2,\bR)$.
Using the on-shell conditions $k+Dk=0$, $k^2=Dk^2=0$, and
fixing $z=i$ (which is invariant is under the $U(1)$) we get
\begin{equation}
\cA_c={1\over 2\pi}{|z-\bar{z}|^{k\cdot Dk}\over (z-\bar{z})}
={1\over 2\pi(z-\bar{z})}=-{i\over 4\pi} \ .
\end{equation}

The scattering amplitude of one closed string and one open string in
the presence of 
a D-brane is
\begin{eqnarray}
\cA_{co}&=&\langle V_L^{(-1,-1)}(k_1,z)V_R^{(-1,-1)}(Dk_1,\bar{z})
V_o^{(0,0)}(k_2,w)\rangle_{D_2}  \nn
&&\langle c(z)c(\bar{z})c(w)\rangle_{D_2}
\langle e^{-\varphi^+(z)} e^{-\varphi^+(\bar{z})}\rangle_{D_2}
\langle e^{-\varphi^-(z)} e^{-\varphi^-(\bar{z})}\rangle_{D_2}\nn
&=&|z-\bar{z}|^{k_1\cdot Dk_1}|z-w|^{2k_1\cdot k_2}
|\bar{z}-w|^{2Dk_1\cdot k_2} \nn
&&2i\left( {k_2{\cal{J}} ^Lk_1\over w-z}+
{k_2{\cal{J}}^LDk_1\over w-\bar{z}}\right)
{(z-w)(\bar{z}-w)\over z-\bar{z}} \ ,
\end{eqnarray}
where we have omitted the fermionic part of the open string vertex
operator, which does not contribute to the amplitude.

The on-shell conditions are $k_1+Dk_1+2k_2=0$, $k_1^2=(Dk_1)^2=k_2^2=0$.
Fixing $z=i$, $w=0$ we get
\begin{equation}
\cA_{co}=i\bigg(k_2{\cal{J}}^L(k_1-Dk_1)\bigg)=
-2i(k_1{\cal{J}}^Lk_2) \ .
\end{equation}
The on-shell conditions imply that $k_1-Dk_1$ and $k_2$
can be both non-zero only for the $(1+1)$-brane.
Thus, only for  the $(1+1)$-brane
$\cA_{co}\ne 0$.

Two $\cA_{co}$ amplitudes can be contracted by
an open string exchange, resulting in an open string channel
of $\cA_{cc}$, the amplitude of two closed string states on a disk.
This amplitude will be computed in the next subsection.
The above result implies that $\cA_{cc}$ in the presence of
$(1+1)$-branes develops
an open string pole.
We will see that this is indeed the case.

\subsubsection{Two Closed Strings}

The net suerconformal ghost number on a disk must be 2
\cite{fms}. Thus,
we take the following matter sector 
vertex operators
\begin{eqnarray}
&&V_L^{(-1,0)}(z;k_1)V_R^{(-1,0)}(\bar{z};k_1)
={1\over4}(k_1J_+^L\Psi(z))(k_1J_+^RD\Psi(\bar{z}))
e^{ik_1\cdot x(z)}e^{ik_1\cdot Dx(\bar{z})},\nn
&&V^{(0,-1)}_L(w;k_2)V_R^{(0,-1)}(\bar{w};k_2)
={1\over4}(k_2J_-^L\Psi(w))(k_2J_-^RD\Psi(\bar{w}))
e^{ik_2\cdot x(w)}e^{ik_2\cdot Dx(\bar{w})} \ .
\end{eqnarray}
The two point function becomes
\begin{eqnarray}
{1\over8}\left({|z-\bar{z}||w-\bar{w}|\over |z-\bar{w}|^2}\right)^s
\left({|z-w|^2\over |z-\bar{w}|^2}\right)^t
\left({2A\over (z-\bar{z})(w-\bar{w})}
-{2B\over |z-w|^2}+{2C\over |z-\bar{w}|^2}\right) \ .
\end{eqnarray}
Here
\begin{equation}
s=k_1\cdot Dk_1={1\over 2}(k_1+Dk_1)^2,\quad
t=k_1\cdot k_2={1\over 2}(k_1+k_2)^2 \ ,
\end{equation}
and
\begin{eqnarray}
A&\!=\!&(k_1J_+^L\eta^{-1}DJ_-^Rk_1)(k_2J_-^L\eta^{-1}DJ_+^Rk_2),\nn
B&\!=\!&(k_1J_+^L\eta^{-1}J_+^Lk_2)(k_1J_+^RD\eta^{-1}DJ_+^Rk_2),\nn
C&\!=\!&(k_1J_+^L\eta^{-1}DJ_+^Rk_2)(k_1J_+^RD\eta^{-1}J_+^Lk_2) \ .
\label{ABC}
\end{eqnarray}
Setting $z=i,w=iy$ with $0<y<1$ and integrating 
over $y$, we obtain the amplitude:
\begin{eqnarray}
{\cal A}_{cc}=-{i\over2}\left(A\,{\Gamma(s-1)\Gamma(t+1)\over\Gamma(s+t)}
+B\,{\Gamma(s)\Gamma(t)\over\Gamma(s+t)}
-C\,{\Gamma(s)\Gamma(t+1)\over\Gamma(s+t+1)}\right).
\end{eqnarray}
In the computation, it was convenient
to use a new variable $x$ defined by
$y=(1-\sqrt{x})/(1+\sqrt{x})$.

Define
\begin{equation}
J_{\pm}=\eta\pm i\cJ,\quad
\widetilde{J}_{\pm}=\eta\pm i\widetilde{\cJ} \ .
\end{equation}
One can check that
\begin{equation}
J_{\pm}\eta^{-1}J_{\pm}=2J_{\pm},\quad
J_{\pm}\eta^{-1}J_{\mp}=0 \ .
\label{jpm}
\end{equation}
Define also
\begin{equation}
p_1=k_1,~~p_2=Dk_1,~~p_3=Dk_2,~~p_4=k_2 \ ,
\label{momentum;p}
\end{equation}
which obey the relations
\begin{equation}
p_1+p_2+p_3+p_4=0,~~p_i^2=0 \ .
\end{equation}
In terms of these, 
\begin{equation}
s=p_1\cdot p_2,~t=p_2\cdot p_3 \ .
\end{equation}
We also use
\begin{equation}
c_{ij}=-ip_i^I\cJ_{IJ}p_j^J \ .
\end{equation}
$c_{ij}$ obey the identities (\ref{id;c}).
In terms of the momenta $p$, the coefficients $A,B,C$ can be written
as
\begin{eqnarray}
A&\!=\!&\left(p_1J_+^L\eta^{-1}DJ_-^RDp_2\right)
\left(p_4J_-^L\eta^{-1}DJ_+^RDp_3\right),\nn
B&\!=\!&2\left(p_1J_+^Lp_4\right)
\left(p_2DJ_+^RD\eta^{-1}DJ_+^RDp_3\right),\nn
C&\!=\!&\left(p_1J_+^L\eta^{-1}DJ_+^RDp_3\right)
\left(p_2DJ_+^RD\eta^{-1}J_+^Lp_4\right) \ .
\end{eqnarray}
Using (\ref{jljr}), they take the form
\begin{itemize}
\item {\bf A-type}
\begin{eqnarray}
A&\!=\!&4\left(p_1J_+^Lp_2\right)\left(p_4J_-^Lp_3\right),\nn
B&\!=\!&4\left(p_1J_+^Lp_4\right)\left(p_2J_-^Lp_3\right),\nn
C&\!=\!&0 \ .
\end{eqnarray}

\item {\bf B-type}
\begin{eqnarray}
A&\!=\!&0,\nn
B&\!=\!&4\left(p_1J_+^Lp_4\right)\left(p_2J_+^Lp_3\right),\nn
C&\!=\!&4\left(p_1J_+^Lp_3\right)\left(p_2J_+^Lp_4\right) \ .
\end{eqnarray}
\end{itemize}

They can be further simplified
using the identities (\ref{id;c})
\begin{itemize}
\item {\bf A-type}
\begin{equation}
{\cal A}_{cc}=2i\,{\Gamma(s-1)\Gamma(t)\over\Gamma(s+t)}
(t-c_{14})(t+c_{23}) \ .
\end{equation}

\item {\bf B-type}
\begin{equation}
{\cal A}_{cc}=0 \ .
\end{equation}
\end{itemize}

Recall that the B-type boundary conditions correspond to
(2+2)-, (2+0)-, (0+2)- and (0+0)-branes in the $\beta$-string, and
(2+1)- and (0+1)-branes in the $\alpha$-string.
The scattering amplitudes in these cases vanish suggesting a
decoupling from the bulk.
The A-type boundary conditions correspond to 
(1+1)-branes in the $\beta$-string and (1+2)- and (1+0)-branes in the
$\alpha$-string.
The amplitudes in these cases contain
infinite number of massive poles
of both open string($s$-) and closed string($t$-) channels.
This suggests that these D-branes do not provide
consistent $N=2$ boundary conditions.

\subsection{One Closed and Two Open Strings}

We insert one closed string vertex operator and two open string
vertex operators on a disk with the appropriate boundary conditions.
The matter sector of the amplitude reads
\begin{equation}
\langle V^{(-1,0)}_o(z_1;k_1)V^{(-1,0)}_o(z_2;k_2)
V^{(0,-1)}_L(z_3;k_3)V^{(0,-1)}_R(\bar{z}_3;k_3)\rangle_{D_2} \ ,
\end{equation}
where $z_1,z_2$ are real.
Using the doubling technique we get
\begin{eqnarray}
\Big\langle 
\left(k_1J^L_+\Psi(z_1)\right)e^{2ik_1\cdot x(z_1)}
\left(k_2J^L_+\Psi(z_2)\right)e^{2ik_2\cdot x(z_2)}
\left(k_3J^L_-\Psi(z_3)\right)e^{ik_3\cdot x(z_3)}
\left(k_3J^R_-D\Psi(\bar{z}_3)\right)e^{ik_1\cdot Dx(\bar{z}_3)}
\Big\rangle \ .\nn
\end{eqnarray}
Using the momentum conservation law
\begin{equation}
2k_1+2k_2+k_3+Dk_3=0 \ ,
\end{equation}
this becomes
\begin{eqnarray}
&&\left({|z_1-z_3|^2|z_2-z_3|^2 \over |z_1-z_2|^2|z_3-\bar{z}_3|^2}\right)^t
\nn
&&\!\!\!\!\!\cdot\left(
{-2\over (z_1-z_3)(z_2-\bar{z}_3)}
\left(k_1J_+^Lk_3\right)\left(k_2J_+^L\eta^{-1}DJ_+^Rk_3\right)
+{2\over (z_1-\bar{z}_3)(z_2-z_3)}
\left(k_1J_+^L\eta^{-1}DJ_+^Rk_3\right)\left(k_2J_+^Lk_3\right)
\right) \ . \nn
\end{eqnarray}
Here we used $t=2k_1\cdot k_3$.
The ghost sector part reads
\begin{eqnarray}
\langle c(z_1)c(z_3)\tilde{c}(\bar{z_3})\rangle_{D_2}
\langle e^{-\varphi^+(z_1)}e^{-\varphi^+(z_2)}\rangle_{D_2}
\langle e^{-\varphi^-(z_3)}e^{-\tilde{\varphi}^-(\bar{z}_3)}\rangle_{D_2} 
={|z_1-z_3|^2\over z_1-z_2} \ .
\end{eqnarray}
Setting $z_1=0$, $z_2=x$ with $-\infty < x < \infty$ and $z_3=i$ and
integrating over $x$, we obtain the
amplitude
\begin{eqnarray}
{\cal A}_{coo}=2i\pi \,{\Gamma(1-2t)\over \Gamma(1-t)^2}\,
\Big[ 
\left(k_1J_+^Lk_3\right)\left(k_2J_+^L\eta^{-1}DJ_+^Rk_3\right)
+\left(k_1J_+^L\eta^{-1}DJ_+^Rk_3\right)\left(k_2J_+^Lk_3\right)
\Big] \ ,
\end{eqnarray}
and we used $2^{1-2t}B({1\over 2}-t,{1\over 2})=
2\pi {\Gamma(1-2t)/ \Gamma(1-t)^2}$. 
Define
\begin{equation}
p_1=2k_1,~~p_2=2k_2,~~p_3=Dk_3,~~p_4=k_3 \ ,
\end{equation}
which obey
\begin{equation}
p_1+p_2+p_3+p_4=0,~~p_i^2=0 \ ,
\end{equation}
then
\begin{equation}
t=p_1\cdot p_4=p_2 \cdot p_3,~~
s=p_1\cdot p_2=p_3 \cdot p_4=-2t,~~
u=p_1\cdot p_3=p_2 \cdot p_4=-s-t=t \ .
\end{equation}
In terms of these, the amplitude takes the form
\begin{eqnarray}
{\cal A}_{ooc}=-i\pi t\,{\Gamma(-2t)\over \Gamma(1-t)^2}\,
\Big[ 
\left(p_1J_+^Lp_4\right)\left(p_2J_+^L\eta^{-1}DJ_+^RDp_3\right)
+\left(p_1J_+^L\eta^{-1}DJ_+^RDp_3\right)\left(p_2J_+^Lp_4\right)
\Big].
\end{eqnarray}
For the A-type boundary conditions, (\ref{jljr}) and the relation
$J_+\eta^{-1}J_-=0$ imply that ${\cal A}_{ooc}=0$.
For the B-type boundary conditions, (\ref{jljr})
and (\ref{id;c}) imply that ${\cal A}_{ooc}=0$.
The 
space-time filling $(2+2)$-brane in the $\beta$-string is
exceptional.
In this case $D=1$ so that
$p_3=p_4$. It thus follows that $s=t=u=0$.
Taking the limit $t\rightarrow 0$ in the amplitude, one obtains
\begin{equation}
{\cal A}_{ooc}=i\pi c_{12}^2 \ ,
\end{equation}
in agreement with the result of \cite{mar;92}.

\subsection{One Open and Two Closed Strings}

We insert two closed string vertex operators and one open string
vertex operator on a disk with the appropriate boundary conditions.
The matter sector of the amplitude reads
\begin{equation}
\langle V^{(-1,0)}_L(z;k_1)V^{(-1,0)}_R(\bar{z};k_1)
V^{(0,-1)}_L(w;k_2)V^{(0,-1)}_R(\bar{w};k_2)V^{(0,0)}_o(x;k_3)\rangle_{D_2}
\ ,
\end{equation}
where $x$ is real.
Using the doubling technique we get
\begin{eqnarray}
{1\over8}&\!\!\!
\langle 
\left(k_1J^L_+\Psi(z)\right)e^{ik_1\cdot x(z)}
\left(k_1J^R_+D\Psi(\bar{z})\right)e^{ik_1\cdot Dx(\bar{z})}
\left(k_2J^L_-\Psi(w)\right)e^{ik_2\cdot x(w)}
\left(k_2J^R_-D\Psi(\bar{w})\right)e^{ik_2\cdot Dx(\bar{w})}\cdot& \nn
&\cdot\left(-k_3{\cal J}^L\partial x(x)+(k_3J^L_+\Psi(x))
(k_3J^L_-\Psi(x))\right)e^{2ik_3\cdot x(x)}
\rangle& \ .
\end{eqnarray}
It is easy to show that this takes the form
\begin{eqnarray}
&&{1\over8}\left(|z-\bar{z}|^{k_1\cdot Dk_1}|w-\bar{w}|^{k_2\cdot Dk_2}
|z-w|^{2k_1\cdot k_2}|z-\bar{w}|^{2k_1\cdot Dk_2}|z-x|^{4k_1\cdot k_3}
|w-x|^{4k_2\cdot k_3}\right)\nn
&&\!\!\!\!\!\cdot\Bigg[
i\left({A\over (z-\bar{z})(w-\bar{w})}-{B\over |z-w|^2}+
{C\over |z-\bar{w}|^2}\right)
\cdot \left({k_3{\cal J}^Lk_1\over x-z}+{k_3{\cal J}^LDk_1\over x-\bar{z}}+
{k_3{\cal J}^Lk_2\over x-w}+{k_3{\cal J}^LDk_2\over x-\bar{w}}\right) \nn
&&+{E\over (z-x)(\bar{z}-x)(w-\bar{w})}+{F\over (z-\bar{z})(w-x)(\bar{w}-x)}
+{G\over (z-x)(w-x)(\bar{z}-\bar{w})} \nn
&&+{H\over (\bar{z}-x)(\bar{w}-x)(z-w)}
+{I\over (z-x)(\bar{w}-x)(\bar{z}-w)}+{J\over (\bar{z}-x)(w-x)(z-\bar{w})}
\Bigg] \ ,
\end{eqnarray}
where $A,B,C$ are given in (\ref{ABC}), and
\begin{eqnarray}
E&\!=\!&-(k_1J_+^L\eta^{-1}J_+^Lk_3)(k_1J_+^RD\eta^{-1}J_-^Lk_3)
        (k_2J_-^L\eta^{-1}DJ_+^Rk_2),\nn
F&\!=\!&(k_1J_+^L\eta^{-1}DJ_-^Rk_1)(k_2J_-^L\eta^{-1}J_-^Lk_3)
        (k_2J_-^RD\eta^{-1}J_+^Lk_3),\nn
G&\!=\!&(k_1J_+^L\eta^{-1}J_+^Lk_3)(k_2J_-^L\eta^{-1}J_-^Lk_3)
        (k_1J_+^RD\eta^{-1}DJ_+^Rk_2),\nn
H&\!=\!&\Big(
        (k_1J_+^RD\eta^{-1}J_+^Lk_3)(k_2J_-^RD\eta^{-1}J_-^Lk_3)
-(k_1J_+^RD\eta^{-1}J_-^Lk_3)(k_2J_-^RD\eta^{-1}J_+^Lk_3)\Big)
        (k_1J_+^L\eta^{-1}J_+^Lk_2),\nn
I&\!=\!&-(k_1J_+^L\eta^{-1}J_+^Lk_3)(k_2J_-^RD\eta^{-1}J_-^Lk_3)
        (k_1J_+^RD\eta^{-1}J_+^Lk_2),\nn
J&\!=\!&-(k_1J_+^RD\eta^{-1}J_+^Lk_3)(k_2J_-^L\eta^{-1}J_-^Lk_3)
        (k_1J_+^L\eta^{-1}DJ_+^Rk_2) \ .
\end{eqnarray}
One can verify that for A-type D-branes $C=I=J=0$, 
while for B-type D-branes $A=E=F=0$.

Setting $z=i$ and $w=iy$ with $0 < y < 1$ and
integrating over $x$ and $y$, we obtain
\begin{eqnarray}
{\cal A}_{cco}&=&-i2^{k_1\cdot Dk_1+k_2\cdot Dk_2-4} \nn
&&\times\int_0^1dy\int_{-\infty}^{\infty}dx \,y^{k_2\cdot Dk_2-1}
(1-y)^{2k_1\cdot k_2+1} (1+y)^{2k_1\cdot Dk_2+1} (x^2+1)^{2k_1\cdot k_3}
(x^2+y^2)^{2k_2\cdot k_3} \nn
&&\!\!\!\!\!\Bigg[-\left({A\over -4y}+{(1-y)^2C-(1+y)^2B\over (1-y^2)^2}\right)
\cdot \left({k_3{\cal J}^L(k_1-Dk_1)\over x^2+1}+
y\,{k_3{\cal J}^L(k_2-Dk_2)\over x^2+y^2}\right)+ \nn
&&+{-i\over2}{E\over (x^2+1)y}+{-i\over2}{F\over x^2+y^2}+
i(G-H){x^2-y\over (x^2+1)(x^2+y^2)(1-y)}+ \nn
&&+i(I-J){x^2+y\over(x^2+1)(x^2+y^2)(1+y)}\Bigg] \ .
\end{eqnarray}
For (2+2)-branes $A=E=F=0$ and $k_i=Dk_i$. Thus, $G=H$ and $I=J$,
implying that  ${\cal A}_{cco}=0$.
Also, for (2+0)-, (0+2)-, (1+0)-, (0+1)- and (0+0)-branes the
on-shell condition $k_3^2=0$ means $k_3=0$ and therefore ${\cal A}_{cco}=0$.  
For (2+1)-branes, the amplitude should not have poles:
it is found that the residues of the open and closed string massless
poles take the form $\cA_{co}\cdot\cA_{coo}$ and
$\cA_{co}\cdot\cA_{ccc}$ respectively, which were shown to vanish.
In order to see explicitly if unphysical massive poles disappear
in the amplitude, 
we have to perform the integration.
We verified that $\cA_{cco}=0$ at least for the case
$k_1\cdot k_3=0$.
We will leave it as an open problem to show this for a generic
value of $k_1\cdot k_3$.

\subsection{Three and Four-Point Amplitudes of Open Strings}

We insert three (four) open string vertex operators on a disk
with appropriate boundary conditions at the boundary.
Open string vertex operators on any type of D-branes
are obtained  by setting  to zero the components of the momentum
transverse to the D-branes
$k_{\bot}$.
This means that the scattering amplitudes are given
by setting $k_{\bot}$ to zero 
in the amplitude of the space-time filling (2+2)-branes \cite{mar;92}:
\begin{equation}
{\cal A}_{ooo}=c_{12},\quad{\cal A}_{oooo}=0 \ .
\end{equation}
The effective action on the worldvolume of any D-brane 
is obtained by a dimensional reduction of 
the effective action of the $(2+2)$-brane.
Using \cite{mar;92}, we see that D-branes
effective actions in 
${\cal N}=2$ strings are given by a dimensional reduction of
self-dual Yang-Mills theory.

Note that apart from the space-time filling (2+2)-branes, 
only the B-type (2+1)-branes in $\alpha$-string allow non-trivial
propagating degrees of freedom on the worldvolume, and
the above three-point amplitude may be non-trivial.
However, it turns out to vanish
for kinematical reasons.
To see this, we notice that an on-shell momentum on the (2+1)-brane
takes  (in the real basis) the form
\begin{equation}
k_i^I=(k_i\cos\theta_i,~k_i\sin\theta_i,~k_i,~0) \ .
\end{equation}
In terms of this parametrization, $\cA_{ooo}$ reads
\begin{equation}
c_{ij}=-ik_ik_j\sin(\theta_i-\theta_j) \ .
\end{equation}
Using the momentum conservation law, we find
\begin{equation}
\cos(\theta_i-\theta_j)=1 \ ,
\end{equation}
and $\cA_{ooo}=0$ for the $(2+1)$-brane.
Thus, we find that lower-dimensional D-branes in $N=2$ strings
allow no non-trivial open string scattering amplitudes.

\subsection{Summary}

Let us summarize the above computations.
We computed several disk amplitudes for strings in the presence of
A-type
and B-type D-branes.
We found that the one-point function of closed strings
is ${\cal A}_{c}= -{i\over 4\pi}$
for all the D-branes.
The two-point function ${\cal A}_{cc}$
vanishes for the B-type D-branes but exhibits infinite number of massive
poles for A-type D-branes, suggesting an inconsistency.

We found that for all D-branes except the $(1+1)$-brane and the
space-time filling $(2+2)$-brane, the open-closed amplitudes, 
${\cal A}_{co}$ and ${\cal A}_{ooc}$,
vanish. 
The disk amplitudes with two or three closed strings and at least one
open string vertex operators vanish for 
the $(0+0)$-, $(2+0)$-, $(0+2)$-, $(1+0)$ and
$(0+1)$-branes. This is
due to the fact that the inserted open string vertex operators can be
chosen to be in the (0,0)-picture, which vanish kinematically by
imposing the on-shell condition.
Thus, for these D-branes all the
open-closed amplitudes on a disk vanish.
It is likely that for these cases all higher genus
amplitudes vanish, since  picture-changing
operators acting on open string vertex operators always yield a vanishing
operator due to the kinematics.

For the (2+2)-, (2+1)-, (1+2)-, (1+1)-branes, 
the open-closed amplitudes could be non-zero. In fact, we found that
for the $(1+1)$-brane ${\cal A}_{co}=-2i(k_1{\cal J}^Lk_2)$, and
for the (2+2)-brane 
${\cal A}_{coo}=i\pi c_{12}^2$.
However, $\cA_{cco}=0$ for the (2+2)-brane, giving no unphysical
open and closed string poles.


We found that for all the D-branes except the
(2+2)-brane, ${\cal A}_{ooo}$ and ${\cal A}_{oooo}$ vanish.
The amplitudes with more than four open string vertices and no closed
string vertex operator should vanish too,
as these are given by a dimensional reduction of the corresponding
amplitudes for the $(2+2)$-brane which vanish \cite{bv}.
As discussed in \cite{bv}, 
the vanishing of $n$-point open string amplitudes
is valid for any order
of string loops.
Only the space-time filling $(2+2)$-brane has a non-trivial
amplitude ${\cal A}_{ooo}=c_{12}$ \cite{mar;92}.

\section{The Effective D-branes Worldvolume Theory}

As discussed in the previous section, the effective action on D-branes
is given by dimensional reduction of self-dual Yang-Mills(SDYM) theory
in (2+2)-dimensions. The self-dual equation reads
\begin{equation}
F_{IJ}={1\over 2}\epsilon_{IJKL}F^{KL} \ ,
\end{equation}
where
\begin{equation}
F_{IJ}=-[D_I,D_J],~~D_I=\partial_I-A_I \ .
\end{equation}
In components,
\begin{equation}
[D_1,D_2]=[D_3,D_4],~~[D_2,D_3]=-[D_1,D_4],~~[D_3,D_1]=-[D_2,D_4] \ .
\end{equation}

\underline{$(2+2)\rightarrow (2+1)$}:

Define $A_{\mu}=(A_1,A_2,A_3),~A_4=\phi$  and regard all the fields
as $x_4$-independent. The self-dual equations become
\begin{equation}
D^{\mu}\phi=-{1\over 2}\epsilon^{\mu\nu\rho}[D_{\nu},D_{\rho}] \ .
\end{equation}
This is the ``Bogomolny equation'' on $\bR^{2,1}$.

\underline{$(2+2)\rightarrow (2+0)$} :

Define $A_3=\phi_3,~A_4=\phi_4$ and regard all the fields as 
$x_3,x_4$-independent. One gets
\begin{equation}
[D_1,D_2]=[\phi_3,\phi_4],~~D_2\phi_3=-D_1\phi_4,~~
D_1\phi_3=D_2\phi_4 \ .
\end{equation}
In terms of the complex scalar defined by $\Phi=\phi_3-i\phi_4$,
the above equations become
\begin{equation}
[D_1,D_2]=-{i\over 2}[\Phi,\Phi^{\dagger}],~~(D_1-iD_2)\Phi=0 \ .
\label{hitchin}
\end{equation}
This is the ``Hitchin'' system. Recall that if we start from SDYM
in $\bR^4$ of signature $(4,0)$, we end up instead of 
(\ref{hitchin}) with
\begin{equation}
[D_1,D_2]=-{i\over 2}[\Phi,\Phi^{\dagger}],~~(D_1+iD_2)\Phi=0 \ .
\end{equation}

\underline{$(2+2)\rightarrow (0+1)$} :

Define $A_1=\phi_1=-\phi^1,~A_2=\phi_2=-\phi^2,~A_3=\phi_3=+\phi^3$
and regard all the fields as $x_1,x_2,x_3$-independent.
One obtains
\begin{equation}
D_4\phi^i={1\over 2}\epsilon^{ijk}[\phi_j,\phi_k] \ .
\end{equation}
This is the ``Nahm equation''.
It is known that the solutions of the Nahm equation are related by
Nahm transformation to those of the Bolomolny equation.
Nahm transformation is a T-duality
that relates (2+1)-branes with (0+1)-branes.

\underline{$(2+2)\rightarrow (0+0)$} :

Define $A_I=\phi_I=\pm\phi^I$ (depending on the signature) and regard all the fields as $x^I$
independent. The self-dual equations become
\begin{equation}
[\phi_I,\phi_J]={1\over 2}\,\epsilon_{IJKL}[\phi^K,\phi^L] \ .
\end{equation}
This is the ``ADHM'' equation.

\section{D-brane Gravitational Backgrounds}

In this section we construct $N=2$ 
D-brane gravitational backgrounds.
The effective action of the $\beta$-string with a source term
coupling to D-branes is 
\begin{equation}
S_{\beta}=\int d^4x\left(
\eta^{i\bar{j}}\,\partial_i\phi\,\partial_{\bar{j}}\phi
+{1\over 3}\,\phi\,\partial_j\partial_{\bar{i}}\phi\,
\epsilon^{ij}\epsilon^{\bar{i}\bar{j}}\,\partial_i\partial_{\bar{j}}\phi
\right)
+\kappa\mu_p\int d^4x \,\phi\,\delta^{4-p}(x_{\bot}) \ ,
\label{ea;beta}
\end{equation}
where $x_{\bot}$ are the Dirichlet directions.
The field equations read
\begin{equation}
-\partial^2\phi
+\partial_1\partial_{\bar{2}}\phi\,\partial_2\partial_{\bar{1}}\phi
-\partial_1\partial_{\bar{1}}\phi\,\partial_2\partial_{\bar{2}}\phi
+{\kappa\mu_p\over 2}\,\delta^{4-p}(x_{\bot})=0 \ .
\end{equation}
The first three terms correspond to the Plebanski equation \cite{ov},
which is the Ricci flatness condition of the target
space K\"ahler manifold. The
scalar field is identified with the deformation of the flat space
K\"ahler potential.

The effective action of the $\alpha$-string with a D-brane source is
\cite{gos}
\begin{equation}
S_{\alpha}=\int d^4x \,
\eta^{i\bar{j}}\,\partial_i\phi\,\partial_{\bar{j}}\phi
+\kappa\mu_p\int d^4x \, \phi\,\delta^{4-p}(x_{\bot}) \ .
\label{ea;alpha}
\end{equation}
The scalar field corresponds to a deformation of a
potential $\Kt$ around a flat background.
\begin{equation}
\Kt=\eta_{1\bar{1}}z^1\bar{z}^{\bar{1}}-\eta_{2\bar{2}}z^2\bar{z}^{\bar{2}}
+\kappa\phi  \ .
\end{equation}
The target space geometry is encoded  $\Kt$ 
\cite{GHR,Hull}
\begin{eqnarray}
&& g_{1\bar{1}}=\partial_1\partial_{\bar{1}}\Kt \ ,~~
g_{2\bar{2}}=-\partial_2\partial_{\bar{2}}\Kt \ ,\nn
&&B_{1\bar{2}}=\partial_1\partial_{\bar{2}}\Kt \ , ~~
B_{\bar{1}2}=\partial_{\bar{1}}\partial_2\Kt \ , \nn
&& e^{2\Phi}=g_{1\bar{1}} \ ,
\label{alpha;geo}
\end{eqnarray}
where $\Phi$ is the dilaton.
Note that in the $N=2$  $\sigma$-model description $z^1$ corresponds to
an $N=2$ chiral superfield while $z^2$ corresponds to
a twisted $N=2$ chiral superfield.

In order to construct the D-branes gravitational backgrounds
we assume
translational invariance along the worldvolume directions and
rotational invariance along the transverse directions.

We first consider D-branes in $\beta$-string.

\underline{(2+2)-brane}: The field equation is
\begin{equation}
-\partial^2\phi
+\partial_1\partial_{\bar{2}}\phi\partial_2\partial_{\bar{1}}\phi
-\partial_1\partial_{\bar{1}}\phi\partial_2\partial_{\bar{2}}\phi
+{\kappa\mu_p\over 2}=0 \ .
\end{equation}
The space-time filling brane has no transverse directions.
If we impose the translational invariance along the world volume
directions, $\phi$ must be a constant.
However, no constant solutions are allowed.

\underline{(2+0)-branes}: Assuming the
translational invariance along the worldvolume directions $z_1$ and
the rotational invariance along the transverse directions, $\phi$
must take the form
\begin{equation}
\phi=\phi(|z_2|) \ .
\end{equation}
The field equation reads 
\begin{equation}
-\partial_2\partial_{\bar{2}}\phi
+{\kappa\mu_p\over 2}\,\delta^{2}(z_2)=0 \ .
\end{equation}
Thus,  $\phi$ is given by the harmonic function in $\bR^2$:
\begin{equation}
\phi={\kappa\mu_2\over 4\pi}\,\log |z_2|^2 \ .
\end{equation}
The K\"ahler potential is given by
$K=-|z_1|^2+|z_2|^2+\kappa\phi$
and the target space metric is 
\begin{equation}
g_{1\bar{1}}=-1,~~g_{2\bar{2}}=1+{\kappa^2\mu_2\over 2}
\delta^2(z_2) \ .
\end{equation}
The Riemann tensor reads
\begin{equation}
R_{i\bar{j}}=\partial_i\partial_{\bar{j}}\log\det g_{i\bar{j}}
=\partial_i\partial_{\bar{j}}\log\left(1+{\kappa^2\mu_2\over 2}
\delta^2(z_2)\right) \ .
\end{equation}
The geometry is
singular at the location of the branes, $z_2=0$.
Note that
\begin{equation}
\phi={\kappa\mu_2\over 4\pi}\,(\log z_2+\log \bar{z}_2) \ ,
\end{equation}
suggesting that $\phi$ is a singular K\"ahler transformation.

\underline{(0+0)-brane}: Consider the 
(4,0) signature. We take 
\begin{equation}
\phi=\phi(r) \ ,
\end{equation}
with $r^2=|z_1|^2+|z_2|^2$.
The interaction terms can not be neglected and 
$\phi$ is not a harmonic function in $\bR^4$.
We will construct a smeared solution later.

Consider next the D-branes of the $\alpha$-string. 
$\phi$ is given by the harmonics function
in $\bR^{4-p}$.

\underline{(2+1)-branes}: We find
\begin{equation}
\phi={\kappa\mu_3\over 2}|x_4| \ .
\end{equation}
Using the formula (\ref{alpha;geo}),
the target space metric and B-fields are given by 
\begin{eqnarray}
&&g_{1\bar{1}}=-1,~~
g_{2\bar{2}}=1+{\kappa^2\mu_3\over2}\delta(x_4) \ , \nn
&&B_{1\bar{2}}=B_{\bar{1}2}=0 \ . 
\end{eqnarray}

\underline{(0+1)-branes}:
As before, we do the analytic continuation. Then
\begin{eqnarray}
&&g_{1\bar{1}}=1+{\kappa^2\mu_1\over 8\pi}
{2x_3^2-x_1^2-x_2^2\over r^5},~~
g_{2\bar{2}}=1+{\kappa^2\mu_1\over 8\pi}
{2x_3^2-x_1^2-x_2^2\over r^5}-{\kappa^2\mu_1\over 2}\delta^3(x_{\bot}), 
\nn
\nn
&&B_{1\bar{2}}=-{3\kappa^2\mu_1\over 8\pi}{(x_1-ix_2)x_3\over r^5},~~
B_{\bar{1}2}=-{3\kappa^2\mu_1\over 8\pi}{(x_1+ix_2)x_3\over r^5} \ .
\end{eqnarray}

Using this background  we can construct 
a (0+0)-branes background by a Legendre transformation \cite{oz;n2}. This
yields a smeared solution
rather than a fully localized one.
The potential of the (0+1)-branes background is given by
\begin{eqnarray}
\Kt&\!=\!&|z_1|^2-|z_2|^2+\phi \nn
&\!\sim\!&|z_1|^2-{1\over 2}(z_2+\bar{z}_2)^2
-{\kappa\mu_1\over 4\pi}{1\over\sqrt{2|z_1|^2+{1\over 2}(z_2+\bar{z}_2)^2}}
\ .
\end{eqnarray}
Here we used a symmetry  transformation that leaves the background
invariant.
The  K\"ahler potential of $\beta$-string is given by the Legendre
transformation:
\begin{equation}
K(z_1,\bar{z}_1,w_2+\bar{w_2})\equiv \Kt-(w_2+\bar{w}_2)(z_2+\bar{z}_2) \ ,
\end{equation}
where $z_2+\bar{z}_2$ can be written as a function of $w_2+\bar{w}_2$
by solving
\begin{equation}
w_2+\bar{w}_2={\partial\Kt\over \partial(z_2+\bar{z}_2)}
=-(z_2+\bar{z}_2)+{\kappa\mu_1\over 8\pi}\,
{z_2+\bar{z}_2\over\left(2|z_1|^2+{1\over 2}(z_2+\bar{z}_2)^2\right)^{3/2}}
\ .
\end{equation}

\section{Discussion}

In this section we would like to discuss some aspects of possible
open/closed $N=2$ string dualities.
A natural approach would be to use the strategy employed for
$N=1$ superstrings. There, one considers 
two descriptions of D-branes: the perturbative description of D-branes
as hyperplanes on which open strings end, and the supergravity
description as solitonic solutions of the field equations.
One then uses the decoupling limit in order to separate
the open string degrees of freedom from the closed string ones
\cite{Maldacena:1997re}.
The decoupling limit for D-branes of $N=1$ superstrings is the low
energy limit $l_s\rightarrow 0$.
One may expect that in our case such a limit is already implemented
since the $N=2$ strings have no massive string states.
Indeed, one support for this seems to be the worldsheet computation of
open and closed strings correlators in section 3.
The vanishing of the correlators without taking any decoupling limit
indicates
that at least perturbatively, 
open and closed strings decouple.

However, upon taking the decoupling limit,
the supergravity backgrounds of
D-branes in $N=1$ superstrings
provide a solution of the low-energy $N=1$ closed string equations.
Here we have solutions of the closed string field equations
 modified by a source term. Thus, without taking a limit
we do not have an $N=2$  closed string background.

We may try to remove the source term by taking another limit.
In this case we are back in a flat
background
without a backreaction of the D-branes.
It is possible that indeed in flat space the D-branes backreaction
is trivial.
It means that we should think about the $N=2$ string as
a topological string
with really non-trivial structure of amplitudes only in curved spaces.

%


In curved spaces one could imagine that the duality would work as
in topological field theories \cite{gv}
or the, recently discussed,
$c=1$ non-critical strings \cite{McGreevy:2003kb}.
One would expect that the effect of D-branes on the curved background
would be to modify some moduli of the $N=2$ closed string background.
Let us see how this can happen.

Consider, for instance,  the $(2+0)$ D-brane.
As shown in section 4, the effective worldvolume theory is the Hitchin system.
Let the curved space be of the form
$T^{\ast}\Sigma_{g}$, where
$\Sigma$ is a genus $g$ Riemann surface.
When the signature is $(4,0)$ $g=0,1$ while
when the signature is $(2,2)$  $g \geq 1$.
Such $(2,2)$ signature solutions have been constructed in \cite{ov}.
The Hitchin system in this case consists of a flat $U(1)$ connection
and an harmonic one-form on $\Sigma_{g}$.
The harmonic one-form can be viewed as an infinitesimal deformation of
$\Sigma_{g}$ to a Riemann surface $\Sigma_{g}'$. 
One can view then the effect of the D-brane as a modification
of the  $T^{\ast}\Sigma_{g}$ background.
The choice of a flat $U(1)$ connection can be mapped
to a deformation of 
$T^{\ast}\Sigma_{g}$ by tensoring the cotangent bundle
by it.
Thus, one may conjecture that $N=2$ closed strings on
 $T^{\ast}\Sigma_{g}$ with a $(2+0)$ D-brane is equivalent
to  $N=2$ closed strings on the deformed  $T^{\ast}\Sigma_{g}$.

Another possible approach to finding a duality could be a flop
transition.
This idea is an extension of the $G_2$ flop \cite{amv},
which was employed to explain the geometric transition \cite{gv}.
Consider as before the (2+0)-branes of the $\beta$-string on 
$T^{\ast}\Sigma_g$, with the 
D-branes wrapping the base $\Sigma_g$.  The D-branes
create a singularity of the background in their position.
We can look for a smooth background
by going smoothly to a negative volume of $\Sigma_g$.
Thus, another possible duality is between $N=2$ closed strings 
in the presence of D-branes and $N=2$ closed strings 
on the background after the flop transition.

We expect that a similar discussion applies to the other D-branes.
It seems interesting to explore such  possible dualities by computing
the topological amplitudes.
The techniques of \cite{ov;n=2}
could be useful.

\section*{Acknowledgements}

We would like to thank Z. Yin for participation in the early stages
of this work, and O. Aharony and D. Kutasov for a valuable discussion.
This research is supported by the US-Israel Binational Science
Foundation.


\newpage

\begin{thebibliography}{99}

\bibitem{ademollo;76}
M.~Ademollo {\it et al.},
``Dual String With U(1) Color Symmetry,''
Nucl.\ Phys.\ B {\bf 111}, 77 (1976).

\bibitem{Gates:1988tn}
S.~J.~Gates, L.~Lu and R.~N.~Oerter,
``Simplified SU(2) Spinning String Superspace Supergravity,''
Phys.\ Lett.\ B {\bf 218} (1989) 33.

\bibitem{oz;n2}
Y.~K.~Cheung, Y.~Oz and Z.~Yin,
``Families of N = 2 strings,''
arXiv:hep-th/0211147.


\bibitem{ov}
H.~Ooguri and C.~Vafa,
``Geometry of N=2 strings,''
Nucl.\ Phys.\ B {\bf 361}, 469 (1991).



\bibitem{gos}
D.~Gluck, Y.~Oz and T.~Sakai,
``The Effective Action and Geometry of Closed N=2 Strings,''
arXiv:hep-th/0304103.

\bibitem{Hull}
C.~M.~Hull,
``The geometry of N = 2 strings with torsion,''
Phys.\ Lett.\ B {\bf 387}, 497 (1996)
[arXiv:hep-th/9606190].


\bibitem{ooy}
H.~Ooguri, Y.~Oz and Z.~Yin,
``D-branes on Calabi-Yau spaces and their mirrors,''
Nucl.\ Phys.\ B {\bf 477}, 407 (1996)
[arXiv:hep-th/9606112].


\bibitem{js;2001}
K.~Junemann and B.~Spendig,
``D-brane scattering of N = 2 strings,''
Phys.\ Lett.\ B {\bf 520}, 163 (2001)
[arXiv:hep-th/0108069].


\bibitem{bogomolny}
E.~B.~Bogomolny,
``Stability Of Classical Solutions,''
Sov.\ J.\ Nucl.\ Phys.\  {\bf 24}, 449 (1976)
[Yad.\ Fiz.\  {\bf 24}, 861 (1976)].


\bibitem{hitchin}
N.~J.~Hitchin,
``The Selfduality Equations On A Riemann Surface,''
Proc.\ Lond.\ Math.\ Soc.\  {\bf 55}, 59 (1987).

\bibitem{Nahm:1979yw}
W.~Nahm,
``A Simple Formalism For The Bps Monopole,''
Phys.\ Lett.\ B {\bf 90} (1980) 413.


\bibitem{adhm}
M.~F.~Atiyah, N.~J.~Hitchin, V.~G.~Drinfeld and Y.~I.~Manin,
``Construction Of Instantons,''
Phys.\ Lett.\ A {\bf 65} (1978) 185.



\bibitem{lz}
U.~Lindstrom and M.~Zabzine,
``N = 2 boundary conditions for non-linear sigma models and  Landau-Ginzburg models,''
JHEP {\bf 0302}, 006 (2003)
[arXiv:hep-th/0209098].



\bibitem{GHR}
S.~J.~Gates, C.~M.~Hull and M.~Rocek,
``Twisted Multiplets And New Supersymmetric Nonlinear Sigma Models,''
Nucl.\ Phys.\ B {\bf 248}, 157 (1984).



\bibitem{fms}
D.~Friedan, E.~J.~Martinec and S.~H.~Shenker,
``Conformal Invariance, Supersymmetry And String Theory,''
Nucl.\ Phys.\ B {\bf 271}, 93 (1986).


\bibitem{mar;92}
N.~Marcus,
``The N=2 open string,''
Nucl.\ Phys.\ B {\bf 387}, 263 (1992)
[arXiv:hep-th/9207024].


\bibitem{hk;review}
A.~Hashimoto and I.~R.~Klebanov,
``Scattering of strings from D-branes,''
Nucl.\ Phys.\ Proc.\ Suppl.\  {\bf 55B}, 118 (1997)
[arXiv:hep-th/9611214].



\bibitem{Polchinski:1995mt}
J.~Polchinski,
``Dirichlet-Branes and Ramond-Ramond Charges,''
Phys.\ Rev.\ Lett.\  {\bf 75} (1995) 4724
[arXiv:hep-th/9510017].



\bibitem{bv}
N.~Berkovits and C.~Vafa,
``N=4 topological strings,''
Nucl.\ Phys.\ B {\bf 433}, 123 (1995)
[arXiv:hep-th/9407190].








\bibitem{Maldacena:1997re}
J.~M.~Maldacena,
``The large N limit of superconformal field theories and supergravity,''
Adv.\ Theor.\ Math.\ Phys.\  {\bf 2} (1998) 231
[Int.\ J.\ Theor.\ Phys.\  {\bf 38} (1999) 1113]
[arXiv:hep-th/9711200].


\bibitem{gv}
R.~Gopakumar and C.~Vafa,
``On the gauge theory/geometry correspondence,''
Adv.\ Theor.\ Math.\ Phys.\  {\bf 3}, 1415 (1999)
[arXiv:hep-th/9811131].


\bibitem{McGreevy:2003kb}
J.~McGreevy and H.~Verlinde,
``Strings from tachyons: The c = 1 matrix reloated,''
arXiv:hep-th/0304224;
J.~McGreevy, J.~Teschner and H.~Verlinde,
``Classical and Quantum D-branes in 2D String Theory,''
arXiv:hep-th/0305194;
I.~R.~Klebanov, J.~Maldacena and N.~Seiberg,
arXiv:hep-th/0305159.



\bibitem{amv}
M.~Atiyah, J.~M.~Maldacena and C.~Vafa,
``An M-theory flop as a large N duality,''
J.\ Math.\ Phys.\  {\bf 42}, 3209 (2001)
[arXiv:hep-th/0011256].










\bibitem{ov;n=2}
H.~Ooguri and C.~Vafa,
``All loop N=2 string amplitudes,''
Nucl.\ Phys.\ B {\bf 451}, 121 (1995)
[arXiv:hep-th/9505183].

\end{thebibliography}
\end{document}